\documentclass[12pt]{emulateapj}
\usepackage{natbib}

\usepackage{apjfonts} % turn on and 'q' through errors to get a nicer font?
\usepackage{comment}

\newcommand{\lir}{L_{\mbox{\scriptsize IR}}}

\newcommand{\eg}{e.g.,}

\newcommand{\mum}{$\mu$m}

\newcommand{\hmpc}{\hbox{$h^{-1}$ Mpc}}

\def\spose#1{\hbox to 0pt{#1\hss}}
\def\simlt{\mathrel{\spose{\lower 3pt\hbox{$\mathchar"218$}}
     \raise 2.0pt\hbox{$\mathchar"13C$}}}
\def\simgt{\mathrel{\spose{\lower 3pt\hbox{$\mathchar"218$}}
     \raise 2.0pt\hbox{$\mathchar"13E$}}}

\slugcomment{ApJ Letters, in press}

\shorttitle{Clustering of Dust--Obscured Galaxies at $z\sim 2$}
\shortauthors{Brodwin et al.}

\begin{document}


\title{Clustering of Dust--Obscured Galaxies at $z\sim 2$}


\author{Mark Brodwin\altaffilmark{1}, 
Arjun Dey\altaffilmark{1},
Michael J.~I.~Brown\altaffilmark{2},
Alexandra Pope\altaffilmark{1}
Lee Armus\altaffilmark{3},
Shane Bussmann\altaffilmark{1},
Vandana Desai\altaffilmark{3},
Buell T.~Jannuzi\altaffilmark{1},
Emeric Le Floc'h\altaffilmark{4}
}


\altaffiltext{1}{NOAO, 950 N.~Cherry Ave., Tucson, AZ 85719}
 \altaffiltext{2}{School of Physics, Monash University, Clayton
  3800, Victoria, Australia}
 \altaffiltext{3}{Spitzer Science Center, Caltech, Pasadena, CA 91125}
 \altaffiltext{4}{Institute for Astronomy, University of Hawaii, Honolulu, HI 96822}

%\altaffiltext{2}{Departments of Physics and Astronomy, UC Berkeley,  CA}


\begin{abstract}

  We present the angular autocorrelation function of 2603
  Dust--Obscured Galaxies (DOGs) in the Bo\"otes field of the NOAO
  Deep Wide-Field Survey.  DOGs are red, obscured galaxies, defined as
  having $R-[24] \ge 14$ ($F_{24}/F_R \ga 1000$).  Spectroscopy
  indicates that they are located at $1.5 \la z \la 2.5$.  We find
  strong clustering, with $r_0 = 7.40^{+1.27}_{-0.84}$ \hmpc\ for the
  full $F_{24} > 0.3$ mJy sample.  The clustering and space density of
  the DOGs are consistent with those of sub-mm galaxies, suggestive of
  a connection between these populations.  We find evidence for
  luminosity-dependent clustering, with the correlation length
  increasing to $r_0 = 12.97^{+4.26}_{-2.64} $ \hmpc\ for brighter
  ($F_{24} > 0.6$ mJy) DOGs.  Bright DOGs also reside in richer
  environments than fainter ones, suggesting these subsamples may not
  be drawn from the same parent population.  The clustering amplitudes
  imply average halo masses of $\log{M} = 12.2 ^{+0.3}_{-0.2}
  \,M_\odot$ for the full DOG sample, rising to $\log{M} = 13.0
  ^{+0.4}_{-0.3} \, M_\odot$ for brighter DOGs.  In a biased structure
  formation scenario, the full DOG sample will, on average, evolve
  into $\sim 3\, L_*$ present-day galaxies, whereas the most luminous
  DOGs may evolve into brightest cluster galaxies.

\end{abstract}


%\keywords{clusters: globular, peanut---bosons: bozos}



\section{Introduction}

The bulk of the stellar mass in the universe is created at $1<z<3$
\citep[\eg][]{dickinson03, rudnick06}.  At $z\approx 1$ this enhanced
star formation occurs primarily in Luminous Infrared Galaxies (LIRGs;
$10^{11} \le \lir\, (L_\odot) < 10^{12}$) \citep{lefloch05}, and by
$z\approx 2$ LIRGs and Ultraluminous Infrared Galaxies (ULIRGs;
$\lir\, (L_\odot) \ge 10^{12}$) dominate the star formation rate (SFR)
budget \citep[\eg][]{caputi07}.  Studies of the spatial distribution
of subsets of the $1<z<3$ ULIRG population with red optical to mid-IR
colors have found very strong clustering \citep{farrah06,
  magliocchetti08}, spanning the range seen from $z\sim 2$ sub-mm
galaxies \citep[SMGs;][]{blain04} to high redshift $z\ga 1$ clusters
\citep{brodwin07}.

\citet[][hereafter D08; see also \citealt{fiore08}]{dey08} presented a
sample of IR-luminous Dust--Obscured Galaxies (DOGs) selected via a
simple optical/mid-IR color cut.  The space density and redshift
distribution of these DOGs are similar to those of sub-mm selected
galaxies \citep{chapman05, coppin06}.  Studies of the spectral energy
distributions (SEDs) of DOG samples show they contain both starburst--
and AGN--dominated galaxies, with the AGN fraction increasing with
luminosity.  Bright ($F_{24} \approx 1$ mJy) DOGs in Bo\"otes have
SEDs of warm AGN ULIRGs \citep{tyler08}, whereas the majority of faint
($F_{24} > 0.1$ mJy; $\left<F_{24}\right>\footnote{Angle brackets
  refer to median values.}  = 0.18$ mJy) DOGs in the GOODS-N field are
dominated by star formation \citep[][hereafter P08]{pope08_dogs}.  In
this Letter, we study the clustering and environment of DOGs as a
function of luminosity to explore their role in galaxy formation at
the key $z\sim 2$ epoch.

We use a concordance cosmology with $\Omega_M = 0.3$ and
$\Omega_\Lambda = 0.7$.  Magnitudes are Vega-relative.  We report
correlation lengths in units of comoving \hmpc, with $H_0 = 100 \,h$
km s$^{-1}$ Mpc$^{-1}$.  All other physical quantities assume $h=0.7$.

\section{Dust--Obscured Galaxies}
\label{Sec: sample}

We study the DOG sample presented by D08.  DOGs were identified in the
Bo\"otes field of the NOAO Deep Wide-Field Survey (NDWFS\footnote{See
  also http://www.noao.edu/noaodeep/}, \citealt{ndwfs99}) via a simple
optical/infrared color selection: $R-[24] > 14$, or equivalently,
$F_{24}/F_R \ga 1000$.  Down to a flux density limit of $F_{24} > 0.3$
mJy ($\approx 6 \sigma$) 2603 sources satisfy this criterion in 8.140
deg$^2$ in Bo\"otes.  Spectroscopy (\citealt{houck05, weedman05,
  brand07, desai08, desai08_irs}) of 86 DOGs indicate that they lie in
a relatively narrow range of redshifts, well-parametrized by a
Gaussian with $\bar{z} = 1.99$ and $\sigma = 0.45$ (D08).

\section{Clustering of DOGs}
\label{section method}

The angular autocorrelation function (ACF) of DOGs is computed as a
function of apparent brightness.  Given the narrow range of redshifts
(Figure 7 of D08), this binning by flux density is to good
approximation a probe of the luminosity dependence of the clustering.

The ACF, parametrized as a simple power law, $\omega(\theta) =
A_{\omega}\theta^{-\delta}$, can be deprojected \citep{limber} to
yield a measurement of the real--space correlation length, $r_0(z)$,
over the redshift range spanned by the 2--D sample:

\begin{equation}
r_0^\gamma(z) = A_{\omega} \left[\frac{H_0 H_{\gamma}}{c}\, \frac{\int_{z_1}^{z_2} N^2(z)\,[x(z)]^{1-\gamma}\, E(z)\, dz}{[\int_{z_1}^{z_2} N(z)\, dz]^2}\right]^{-1}.
\label{Eq: r_0}
\end{equation}

Here $\gamma \equiv 1+ \delta$, $H_{\gamma} =
\Gamma(1/2)\,\,\Gamma[(\gamma-1)/2]/\Gamma(\gamma/2)$, $N(z)$ is the
redshift distribution, and $E(z)$ and $x(z)$ describe the evolution of
the Hubble parameter and the comoving radial distance, respectively.
The primary uncertainty in the inferred real-space correlation length
comes from uncertainty in the shape of the redshift distribution.

We calculate the ACF using the \citet{hamilton} estimator:
\begin{equation}
\omega_{\mbox{\scriptsize H}} = \frac{DD\times RR }{DR\times DR} - 1
\label{Eq: hamilton}
\end{equation}
Here $DD$, $DR$ and $RR$ are the sum of ordered data--data,
data--random, and random--random pairs at each angular separation.  We
used 500,000 randoms to ensure a robust Monte Carlo integration.  We
also computed the ACF using the \citet{landy&szalay} estimator and
find nearly identical results.

Regions of the survey affected by cosmetic artifacts or data quality
issues can compromise a robust measurement of clustering.  Masking is
implemented in the images and random catalogs to reject these areas.

\begin{figure}[bthp]
\epsscale{1.}
\plotone{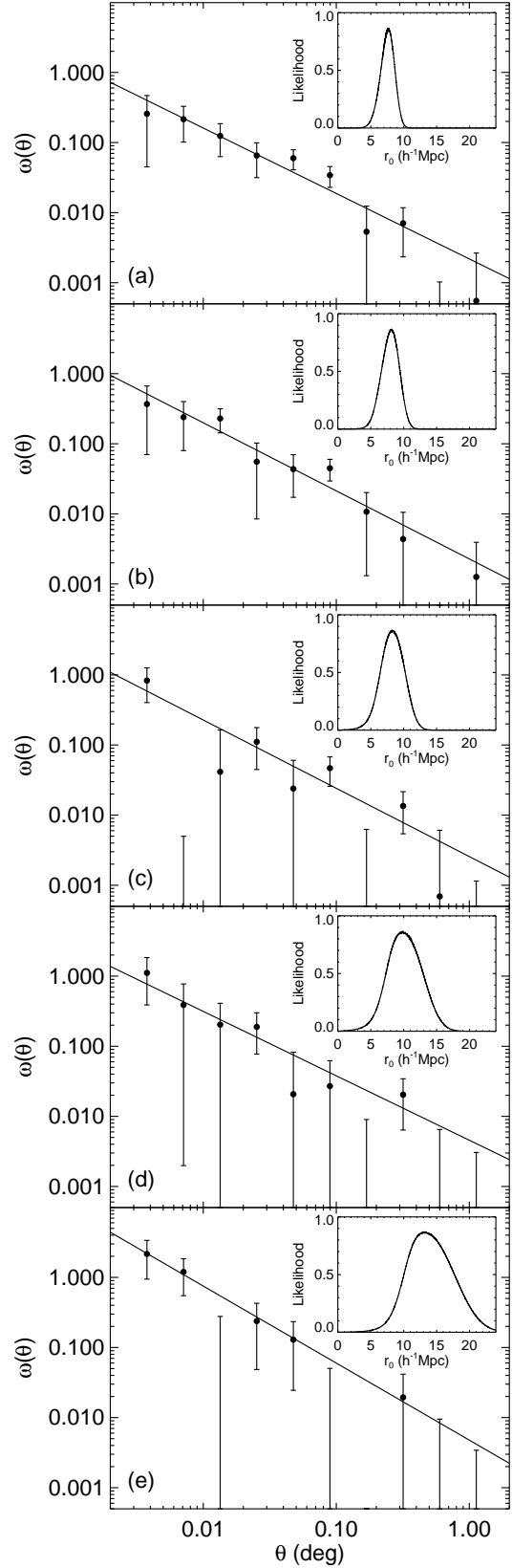}
\epsscale{1}
\caption{Angular correlation functions for (a) the full DOG sample,
  (b) fainter DOGs with $0.3 \le F_{24}$ (mJy) $ < 0.5$, and brighter
  DOGs with (c) $F_{24} > 0.4$, (d) $F_{24} > 0.5$, and (e) $F_{24} >
  0.6$ mJy.  The red lines show the best $\chi^2$ fits; the insets show
  the Bayesian likelihood functions in $r_0$.  All slopes are
  consistent with $\delta = 0.9$.}
\label{Fig: ACF}
\end{figure}

A Bayesian technique is used to determine the correlation lengths.
This allows marginalization over the slope, $\delta$, subject to the
weak prior that $0.2 \le \delta \le 1.8$. This is desirable since the
small size of some of the subsamples precludes precise simultaneous
measurements of both the amplitude and the slope.  We show in Figure
\ref{Fig: ACF} the simple $\chi^2$ fits and the Bayesian likelihood
functions in $r_0$ computed using the redshift distribution from D08.
The slopes in the $\chi^2$ fits for all samples are consistent with
$\delta = 0.9$.  Results are summarized in Table \ref{Tab: Fits}.

\begin{deluxetable}{ccrrrrr}
\tablecolumns{6}
\tabletypesize{\normalsize}
\tablecaption{Clustering of DOGs in Bo\"otes\label{Tab: Fits}}
\tablewidth{0pt}
\tablehead{
\colhead{$F_{24}$} & \colhead{$\left<F_{24}\right>$} & & \multicolumn{1}{c}{$r_0$\tablenotemark{a}} &  & \multicolumn{1}{c}{Halo Mass} &\\
\colhead{(mJy)} & \colhead{(mJy)} & \multicolumn{1}{c}{$N$}& \multicolumn{1}{c}{(\hmpc)} & \multicolumn{1}{c}{$b$} & \multicolumn{1}{c}{$\log (M/M_\odot)$} & \multicolumn{1}{c}{$L/L_* (z=0)$}
}
\startdata
$>0.3$       & 0.40 & 2603 &  $7.40 \pm ^{1.27}_{0.84}$ & $ 3.12 \pm^{0.51}_{0.34}  $ &$  12.2\pm ^{0.3}_{0.2}$ &$  3.4\pm ^{0.5}_{0.7}$ \\
$0.3$--$0.5$ & 0.36 & 1846 &  $7.99 \pm ^{1.41}_{1.30}$ & $ 3.36 \pm^{0.56}_{0.52}  $ &$  12.3\pm ^{0.2}_{0.3}$ &$  3.7\pm ^{0.8}_{0.8}$ \\
$>0.4$       & 0.53 & 1285 &  $8.66 \pm ^{1.41}_{2.10}$ & $ 3.63 \pm^{0.56}_{0.84}  $ &$  12.5\pm ^{0.2}_{0.4}$ &$  4.1\pm ^{1.2}_{0.8}$ \\
$>0.5$       & 0.65 &  757 & $10.19 \pm ^{2.47}_{2.64}$ & $ 4.24 \pm^{0.97}_{1.05}  $ &$  12.7\pm ^{0.3}_{0.5}$ &$  5.0\pm ^{1.5}_{1.4}$ \\
$>0.6$       & 0.85 &  454 & $12.97 \pm ^{4.26}_{2.64}$ & $ 5.33 \pm^{1.65}_{1.04}  $ &$  13.0\pm ^{0.4}_{0.3}$ &$  6.6\pm ^{1.5}_{2.4}$ \\
\enddata 
\tablenotetext{a}{Correlation length from the Bayesian fit.  The
  uncertainty range corresponds to the 68\% confidence interval.}
\tablecomments{\small $F_{24}$ is the 24$\mu$m flux density, $N$
  refers to the total numbers of sources in each flux bin, and $b$ is
  the linear bias.  Masses and luminosities assume $h=0.7$.}
\end{deluxetable}

The ACF errors were derived using the full covariance matrix, computed
using the \citet{brown08} implementation of the
\citet{eisenstein_zaldarriaga01} analytic approximation.
Conservatively assuming 5\% of the DOG sample is spurious and
uncorrelated, then at most we are underestimating the clustering by
$\approx$ 11\%.

\begin{figure}[bthp]
\plotone{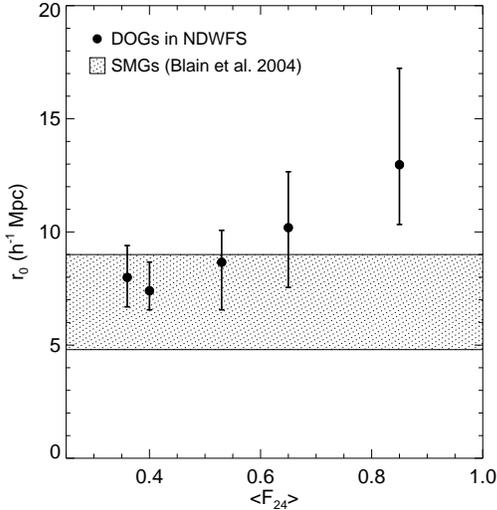}
\caption{Dependence of correlation length on median flux for the
  samples shown in Figure 1 and summarized in Table 1, illustrating
  that clustering strength increases with median luminosity. The
  shaded region shows the 1 $\sigma$ range for SMGs \citep{blain04}.}
\label{Fig: r0 vs flux}
\end{figure}

We show in Figure \ref{Fig: r0 vs flux} the correlation length vs.\
median flux density for several DOG subsamples.  For comparison we
also show the 1 $\sigma$ range for the \citet{blain04} SMG sample.
The trend is strongly suggestive of luminosity-dependent DOG
clustering which, as a consequence, implies that bright and faint DOGs
must reside in different environments.  We show below
(\textsection\ref{Sec: discussion}) that this is observed. While the
uncertainties preclude ruling out the null hypothesis of no luminosity
dependence from the clustering measurements alone, the clustering and
environmental studies, taken together, provide convincing evidence of
an intrinsic difference between bright and faint DOGs.

The primary uncertainty in Limber inversion is the sample redshift
distribution.  If the assumed distribution is broader than the true
distribution the resulting amplitudes will be biased high.  Our DOG
redshift distribution is based on 86 spectroscopic redshifts, drawn
largely from the bright end of the DOG population, and may not be
representative of the fainter DOGs.  If the fainter DOGs are largely
star-formation dominated (\eg\ D08; P08) they would likely have a
narrower redshift distribution due to the strong 7.7$\mu$m PAH
emission feature passing through the 24um filter at $z=2$.  In this
case the evidence for luminosity-dependent clustering would be {\em
  strengthened} since the correct correlation length for fainter DOGs
would be reduced relative to that of the brighter ones.
Quantitatively, if the true redshift distribution were 10\% (30\%)
narrower the correlation lengths would be reduced by 5\% (17\%).

\section{Discussion \& Conclusions}
\label{Sec: discussion}

To similar flux limits ($F_{24} > 0.4$ mJy) the clustering amplitudes
measured for other $z\sim 2$ ULIRG samples are higher than those for
DOGs in Bo\"otes.  In the SWIRE \citep{lonsdale03} survey fields
\citet{farrah06} find correlation lengths of $r_0 = 9.4 \pm 2.24
\hmpc$ and $r_0=14.4\pm 1.99 \hmpc$ for ULIRG samples at $1.5 <z< 2$
and $2<z<3$, respectively.  In a 0.7 deg$^2$ subset of the same survey
\citet{magliocchetti08} find $r_0 = 15.9 ^{+2.9}_{-3.4} \hmpc$ for a
sample of 210 $z\sim 2$ ULIRGs.  Beyond the fact that the samples in
these works have different selection criteria, from each other and
from the DOG sample, the most likely explanation for the differences
with our work is that they adopt broader redshift distributions based
on photometric redshifts.  \citet{desai08} and D08 demonstrate that
optical/IR photometric redshifts can be unreliable for these heavily
obscured sources, particularly when the optical detections are
marginal or non-existent.  These previous analyses have likely
underestimated their uncertainties, because they fixed the slope of
the correlation function, eliminating the covariance between the slope
and $r_0$, and adopted simple Poisson errors, ignoring correlations
between adjacent bins.

While DOGs are a mixed population, consisting of both starbursting
galaxies and AGN, a majority are dominated by star formation (P08).
It is interesting then that the space densities (D08) and clustering
are quite similar to SMGs \citep{coppin06,blain04}, which are known to
be star-formation dominated \citep{pope08}. 

On the other hand, the AGN fraction increases with luminosity
\citep[P08; D08;][]{tyler08}.  The very strong clustering of brighter
DOGs implies they are located in rare, rich environments.  Indeed,
\citet{galametz08} find a strong increase in the incidence of AGN in
rich galaxy clusters at $z>1$.  We show in Figure \ref{Fig:
  environments} the surface density profiles of 4.5\mum\ selected
galaxies from the IRAC Shallow Survey
\citep{eisenhardt04,brodwin06_ISS} around DOG samples with several
flux limits.  We only consider IRAC galaxies with colors redder than
$[3.6] - [4.5] > 0.6$, a criterion that selects objects at $z\ga1.5$
\citep{stern05,papovich08}.  Following \citet{padmanabhan08} the mean
space density of these 4.5\mum\ sources has been subtracted.  All DOG
samples are clearly correlated with the red IRAC galaxies, showing a
large excess on small scales.  The mean surface densities of IRAC
sources in the vicinity of the DOGs increases monotonically with their
brightnesses, indicating that brighter DOGs do in fact reside in
richer environments, as suggested by their stronger clustering.  The
environment of the DOGs is particularly rich on small ($\la 250$ kpc)
scales, suggesting that they preferentially reside in groups of
galaxies.  This inference is supported by their large clustering
amplitudes, as well as by recent theoretical work
\citep[e.g.][]{hopkins08} showing that at $z\sim2$ the maximal merging
efficiency of gas-rich halos, and hence resultant starburst activity,
occurs in group-mass halos.

\begin{figure}[bthp]
\plotone{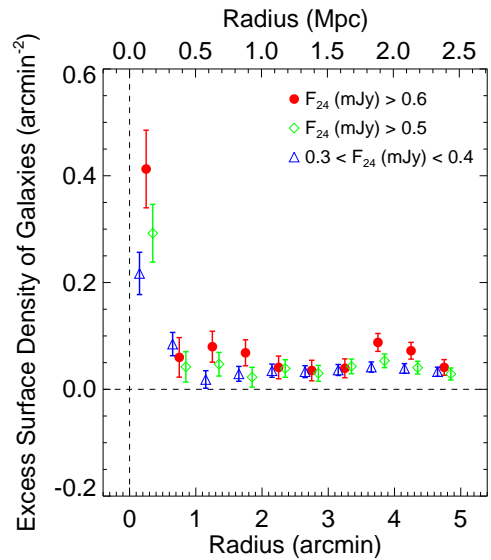}
\caption{Surface profile of $z\ga1.5$ 4.5\mum\ sources in Bo\"otes
  around DOG samples with various flux limits.  The contribution from
  the mean has been subtracted.  There is a clear luminosity
  dependence to the profiles such that brighter DOGs lie in
  increasingly rich environments.  The top axis is in physical
  separations at $z=2$ for our chosen cosmology.  The region within
  $\la 250$ kpc of the DOGs is particularly rich.  Some of the symbols
  are offset slightly in radius for clarity.  }
\label{Fig: environments}
\end{figure}

D08 proposed that both types of DOGs are drawn from the same parent
population, and that the luminosity dependence of the AGN fraction
arises from the additional energy output from those DOGs undergoing an
active AGN phase.  The present results suggest an alternative
explanation.  Although the evidence for luminosity-dependent
clustering is marginal given the large errors, the corroborating
observation that brighter DOGs reside in richer environments than
fainter ones indicates that they are not drawn from identical parent
populations.  This conclusion is robust to uncertainties in the
redshift distribution, provided a single distribution is used for both
bright and faint DOG samples.  If the redshift distribution of faint
DOGs were narrower, the likeliest situation, the strength of the
luminosity dependence would increase.

The linear biases of the DOG samples, listed in Table 1, are computed
as the square root of the ratio of the $z=2$ DOG and dark matter
correlation functions at a scale of $5$ \hmpc, where the latter is
computed following the HaloFit prescription of \citet{smith03}.  The
full sample has a bias of $b=3.12^{+0.51}_{-0.34}$, where the
uncertainty is propagated from the error in the clustering.  The bias
increases with DOG flux, from $b=3.36^{+0.56}_{-0.52}$ for faint DOGs
($\left< F_{24} \right> = 0.36$ mJy) to $b=5.33^{+1.65}_{-1.04}$ for
bright DOGs ($\left< F_{24} \right> = 0.85$ mJy).  Comparison of the
observed clustering with that of halos in a large, high-resolution
numerical simulation \citep[described in detail in][]{brown08}
indicates that the full DOG sample has an average halo mass of
$\log{M} = 12.2 ^{+0.3}_{-0.2} \,M_\odot$.  The masses increase with
DOG luminosity, as shown in Table 1, reaching $\log{M} = 13.0
^{+0.4}_{-0.3} \, M_\odot$ for the brightest DOGs.  In the
\citet{fry96} biased structure formation model, assuming merger--free
passive evolution, these samples will evolve into $\approx 3.4$ and
$\approx 6.6\, L_*$ galaxies by the present day.  The latter have the
masses of brightest cluster galaxies in local clusters.

At low redshift the space density and color bimodality of galaxies can
be modeled by truncating star formation in galaxies above a particular
host halo mass \citep[\eg][]{kauffmann03, croton06, brown08}. Several
papers suggest that the transition halo mass is $\sim 10^{12}$
M$_\odot$, and this mass undergoes negligible evolution at $z<1$
\citep[\eg][]{dekel06, cattaneo08, brown08}. The $z \sim2$ DOGs, most
of which are undergoing vigorous star formation, reside in $\approx
10^{12} - 10^{13}$ M$_\odot$ halos.  Perhaps the mode of gas accretion
onto massive halos changes at $z>1$, as suggested by the virial shock
heating model of \citet{dekel06}.

In summary, the DOG sample presented in D08 is a highly clustered
population of luminous, obscured galaxies at $z\approx 2$, with $r_0 =
7.40^{+1.27}_{-0.84}$ \hmpc.  Their clustering, space density, and
redshift distribution are quite similar to SMGs, indicating that they
reside in similar mass halos and suggesting a possible connection
between these populations.  The clustering strength increases with
luminosity, up to $r_0 = 12.97^{+4.26}_{-2.64}$ \hmpc\ for F$_{24} >
0.6 $ mJy DOGs.  Luminous DOGs also reside in richer environments than
fainter ones.  These results suggest that luminous DOGs, which are
more likely to host active AGNs, are not drawn from the same parent
population as faint ones, but rather reside in more massive halos.
DOGs are highly biased, with $3.1< b< 5.3$, corresponding to masses of
$12.2 < \log{(M/M_\odot)} < 13.0$ over the luminosity range studied
here.  They are a population of vigorously star forming galaxies with
halo masses larger than $10^{12} M_\odot$, the suggested critical mass
for the truncation of star formation. They will likely evolve into
very massive ($3 \la L/L_* \la 7$) local galaxies.  \acknowledgements

MB is grateful to M.~Dickinson and NOAO for supporting this research.
We thank M.~White for providing his HaloFit code, halo correlation
functions, and for a careful reading of the manuscript.  We thank the
anonymous referee for helpful comments.  This work is based in part on
observations made with the {\it Spitzer Space Telescope}, which is
operated by the Jet Propulsion Laboratory, California Institute of
Technology, under a contract with NASA.  We used data products from
the NDWFS, which was supported by NOAO, AURA, Inc., and the NSF.  NOAO
is operated by AURA, Inc., under a cooperative agreement with the
National Science Foundation.

\bibliographystyle{astron3}
\bibliography{bibfile}

\end{document}